\documentclass{v23windows}

\def\Journal#1#2#3#4{{#1} {\bf #2}, #3 (#4)}

\def\be{\begin{equation}}
\def\ee{\end{equation}}
\def\bea{\begin{eqnarray}}
\def\eea{\end{eqnarray}}

\newcommand{\Photo}{\includegraphics[height=35mm]{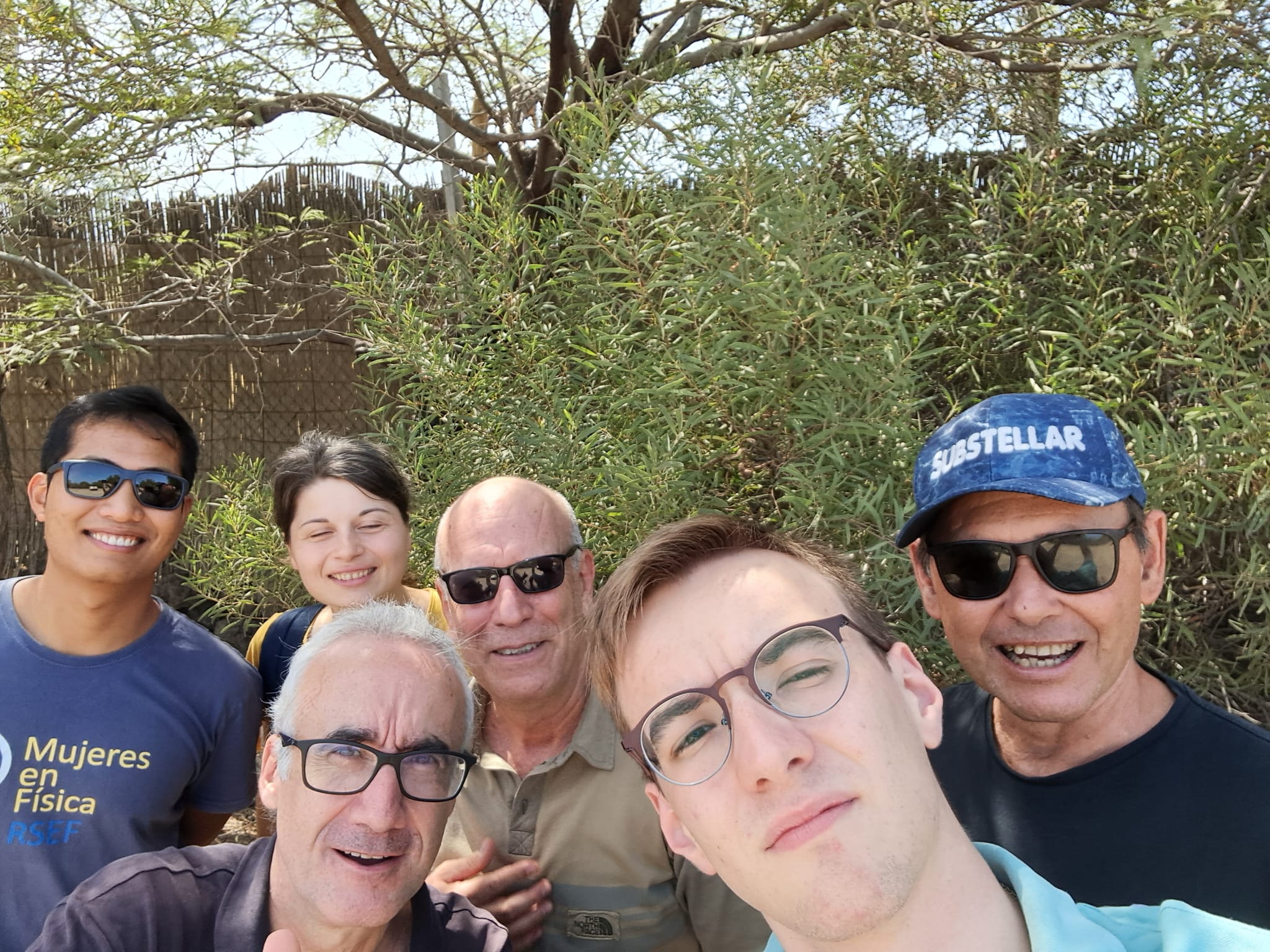}}

\begin{document}
\vspace*{4cm}
\title{Substellar science in the wake of the ESA Euclid space mission}

\author{ Eduardo L. Martín$^1$, Hervé Bouy$^2$, Diego Martín$^1$, Marusa Zerjal$^1$, Jerry J.-Y. Zhang$^1$, Adam Burgasser$^3$, Javier Olivares$^4$, Nicolas Lodieu$^1$, Enrique Solano$^5$, Patricia Cruz$^5$, David Barrado$^5$, Nuria Huélamo$^5$, Pedro Mas-Buitrago$^5$, Maria Morales$^5$, Carlos del Burgo$^1$, Alberto Escobar$^1$, Víctor Sánchez Béjar$^1$, Johannes Sahlmann$^6$ and Maria Rosa Zapatero Osorio$^5$}

\address{$^1$Instituto de Astrofísica de Canarias, San Cristóbal de La Laguna, Tenerife, Spain; $^2$LAB, Université de Bordeaux, France; $^3$ Dept. of Astronomy and Astrophysics, University of California at San Diego, La Jolla, USA; $^4$ Dept. de Inteligencia Artificial, UNED-ETSI, Madrid, Spain; $^5$ INTA-CSIC Centro de Astrobiología, Madrid, Spain; $^6$ ESAC, Madrid, Spain}

\maketitle\abstracts{
 The ESA space mission Euclid was launched on July 1st, 2023 and is undergoing its science verification phase. In this invited review we show that Euclid means a ‘before’ and an ‘after’ for our understanding of ultra-cool dwarfs and substellar-mass objects and their connections with stars, exoplanets and the Milky Way. Euclid enables the study with unprecedented statistical significance a very large ensemble of ultracool dwarfs, the identification of new types of substellar objects, and the determination of the substellar binary fraction and the Initial Mass Function (IMF) in diverse galactic environments from the nearest stellar nurseries to the ancient relics of Galactic formation.}

\section{Background}
 
Substellar-mass objects (SMOs) include brown dwarfs (BDs), free-floating planetary-mass objects (PMOs) and exoplanets that overlap in some of their basic parameters such as the effective temperature and luminosity, and that have in common that they do not have enough mass to sustain stable hydrogen thermonuclear fusion. In this work we avoid to define hard boundaries between BDs, PMOs, and exoplanets because they are likely to evolve as our understanding of SMOs gets more mature. For early views on this topic see the contributions to IAU Symposium 211 \cite{boss}. The term ultra-cool dwarf (UCD) refers to objects cooler than about 2800 K, spectral type late-M ($\ge$M6), L, T or Y. Late-M and L-type UCDs may be SMO or not depending on age and metallicity. T-type and Y-type UCDs are thought to be SMOs for all ages and metallicities and in terms of their effective temperature they are referred here as extremely ultra-cool (EUCDs) dwarfs. 
 
The first bona-fide examples of SMOs were discovered in the last decade of the 20th century. The turning point was brought about by the discovery of three benchmark objects, namely, Teide Pleiades 1, the first unambiguous ultra-cool BD in an open cluster \cite{rebolo95}; 51 Peg b, the first Jupiter-mass exoplanet found around a solar-type star \cite{mayor95}; and Gl 229 B, the first directly-imaged EUCD BD companion of a red dwarf \cite{nakajima95}, with a spectrum very similar to that of Jupiter \cite{oppenheimer95}. Furthermore, the first PMOs with spectroscopic characterization were identified in the Orion star-forming region \cite{lucas00,zapatero00} and the substellar Initial Mass Function (IMF) in the Sigma Orionis Open cluster down to the planetary-mass domain  \cite{bejar00} indicated that substellar objects are very numerous and that there is no apparent cutoff off down to a few times the mass of Jupiter. 

Investigation of the basic properties of SMOs have led to adding new letters to the classic OBAFGKM spectral sequence. First, objects in the range of Teff between 2200K and 1200 K were identified mainly thanks to the DENIS and 2MASS surveys and follow-up optical spectroscopy led to the definition of the L-type class \cite{martin97,martin98,martin99,martin05,martin10,kirkpatrick99,kirkpatrick00}.  L dwarfs have dusty atmospheres and are constituted of a mix of middle-aged BDs, old very-low mass (VLM) stars at the bottom of the main-sequence and young PMOs. Lithium is a very useful probe of their interiors as it provides rather accurate constraints on their ages and masses.  This has been recently demonstrated in the nearest open cluster, the Hyades, where Li abundances have been obtained for the first time in their L-type members\cite{lodieu18,martin18}, and in VLM binaries with dynamical masses measured from orbital motions \cite{martin22}. Cooler than the L dwarfs, the T dwarfs cover the Teff range from 1200 K to 500 K \cite{burgasser06,kirkpatrick19}.  They are thought to include only SMOs. T dwarfs in the halo of the Galaxy are not known yet, although the first examples of metal-poor (iron abundance a factor of 10 or lower than the Sun) candidates have recently been identified using the WISE space mission \cite{schneider20,meisner21}, and follow-up ground-based observations \cite{zhang23}. Halo T dwarfs are within the reach of Euclid, and follow-up on their properties may shed light on the cosmological lithium problem, which is the discrepancy of about a factor of four between the predictions from Big Bang nucleosynthesis and the observational Spite plateau \cite{fields11}. Even cooler than the T-type, the Y-type dwarfs reach down to Teff$\sim$250 K \cite{luhman14} and presumably masses of only a few Jupiters. Dozens of Y dwarfs have been identified within 20 pc from the Sun (see Y dwarf catalog curated and maintained by Mike Cushing). Some of their atmospheres may be suitable for sustaining a significant amount of biota \cite{yates17} that might lead to detectable bioluminiscence \cite{villa}.   

\section{Prospects with Euclid}

 The number of resolved UCDs has grown into the few thousands, thanks to deep and wide-area optical and infrared surveys such as CFBDS, DENIS, DES, 2MASS, HSC, SDSS, PanStarrs, UKIDSS, VHS, and WISE \cite{carnero19}.  Euclid, complemented by other surveys (Vera Rubin LSST, Nancy Roman Space Telescope), will bring over an order of magnitude boost in the numbers of known UCDs (Figure 1) and will enable statistical analysis of different SMO populations.  

Euclid has two instruments, namely the Near Infrared Spectrometer and Photometer (NISP) and the Visual instrument (VIS) \cite{laureijs2011}. They are very stable and always observe in the same mode. They will provide unprecedented deep, wide, homogeneous and high spatial resolution optical and near-infrared imaging and low-resolution spectroscopic surveys over a third of the sky \cite{scaramella22}. Simulations of Euclid performance for UCDs have been presented for photometric detections \cite{solano21} as well as for slitless spectra \cite{martin21}. They have used updated filter profiles \cite{schirmer22} that are available at the Spanish Virtual Observatory (SVO).  
The diversity of UCDs is so rich that the statistics of the different types of objects need large samples. 

\subsection{Substellar multiplicity, and planets around substellar primaries with Euclid}

There are many parameters that may enter in shaping the properties of SMO objects, such as age, angular momentum, binarity, chemical composition, formation environment, rotation and mass. For example, the occurrence rate of SMO companions has been estimated at about 2\% in wide orbits (semimajor axis = a $>$ 100 AU) from direct near-infrared imaging of young stars and BDs \cite{chinchilla20} and $<$ 5\% in close orbits (a $<$ 5 AU) from radial-velocity surveys \cite{grandjean21}. 

The high spatial resolution of the Euclid/VIS (0.1 arcsec) and NISP (0.3 arcsec) images combined with  customized analysis of the point spread function (PSF) using benchmark UCDs will enable optimization of the detection of companions around different types of stars and SMOs by direct imaging.  Previous experience using the Hubble Space Telescope (HST) indicate that VLM binaries can be resolved down to about 1/5th of the spatial resolution \cite{bouy03,martin03}. Custom-made PSF analysis based on UCD templates observed with the stable instrumental configuration of Euclid will allow us to resolve thousands of VLM binaries, probably increasing by a factor of $>$10 the number of these systems known.  

Very faint companions detected by Euclid around young stars and SMOs will be candidate planetary-mass companions in very wide orbits. Around 10$^3$ such objects are expected, increasing by about a factor of $>$10 the number of these systems known.  The approximate range of companion separations covered by Euclid will be from 10 AU to 10,000 AU, and mass ratios from 0.4 to 0.05. Low-gravity spectral features will be analyzed in the Euclid slitless spectra, and they will be used to identify new low-gravity SMOs in the Euclid data.  

For a subset of the objects time-domain astrometric (including proper motions and parallaxes) and photometric information will be obtained. We estimate this subset will consist of around 20,000 L and T dwarfs, 2500 L subdwarfs, 300 T subdwarfs, and dozens of Y dwarfs that will be observed repeatedly by Euclid because they are located in the deep fields.  This information will be used to derive parallaxes and proper motions, and search for weather in VLM stars, UCDs and SMOs, possibly coming from variable clouds \cite{samra20}. Other exotic non-thermal phenomena that we will search for in the Euclid VIS images are auroral emission \cite{vidotto19} and magnetic flares.

Preliminary simulations of the Euclid´s astrometric accuracy suggest enough sensitivity to detect SMO companions with masses between 0.5 and 50 Jupiters with orbital periods $<$ 3 years (1 AU semi-major axis) orbiting a 70 Jupiter-mass primary. The expected astrometric VLM binary yield estimated from an ESO/VLT survey of 20 field L dwarfs \cite{sahlmann14} is that for 1\% of the UCDs located in the Euclid deep fields dynamical masses within the lifetime of the nominal Euclid surveys (6 years) can be measured. This number could be increased with complementary  follow-up observations with telescopes for which have already demonstrated high enough astrometric precision in UCDs (GTC \& VLT). Giant planets around VLM primaries may have lower frequency than binaries. They challenge planet formation models. Interesting planetary candidates will be subject to intensive follow-up efforts. Euclid-based results for planet search around VLM primaries will be compared to other planet searches to increase the understanding of planet formation across a broad range of host star masses.   

\subsection{The substellar IMF with Euclid}

The Initial Mass Function (IMF) represents the distribution of primary masses that emerges from the star-formation process. There are four open problems about the IMF: (1) Is the IMF invariant over the history of the Milky Way? (2) What is the dominant shape of the IMF? (3) Where is the low-mass cutoff of the IMF? (4) What is the effect of metallicity on the IMF? 

Two main IMF shapes have been used in the literature; a power law \cite{salpeter55}, that could have more than one slope \cite{kroupa03}, or a log-normal function \cite{adams96}. The power law indicates that the IMF is mainly defined by a single dominating process, whereas the log-normal suggests that a variety of processes play roles of similar weight. Isothermal MHD turbulence is one of the leading theorerical frameworks \cite{haug18}, but it struggles to make a high enough number of SMOs. Metallicity may play an important role in shaping the IMF \cite{sharda22}. 

Murky SMOs lurk in our immediate neighborhood.  The closest BD known is a resolved binary at just 2.02 pc \cite{luhman13}.  About 70,000 years ago, a binary system made of a VLM star and a BD approached the Solar System at just 0.25 pc, intersecting the Oort cloud \cite{mamajek15}. These findings, together with those reported in star-forming regions and young open clusters and microlensing surveys suggest that there is no clear low-mass cutoff of the IMF or that two different stellar and substellar IMFs overlap in the BD regime \cite{thies15}. A fragmentation low-mass limit to the IMF has been calculated  at a mass of around 10 Jupiters \cite{mondal19}, but it remains to be found observationally. The Euclid Early Release Observation program in nearby star-forming regions i(ERO002, PI E. L. Martín) will showcase the power of Euclid to probe the substellar population of nearby stellar nurseries such as Orion, Perseus and Taurus. An example of the Euclid advantage is shown in Figure 2. 

Open clusters (e.g., IC2391, Hyades and Praesepe) appear to have lost some of their SMO population due to dynamical evolution \cite{wang11}. This could be a source of PMOs (aka rogue planets) wandering in the Milky Way. It has been pointed out that N-body interactions can also lead to a population of SMOs ejected from newly formed systems \cite{vorobyov17}. A possible test for different formation scenarios could be a strong dependence of the binary fraction of SMOs with primary mass \cite{marks17}.  

The conversion from basic parameters to fundamental properties will be made using evolutionary models for VLM stars and SMOs. The models will be tested with the properties of benchmark objects. For example, in the Hyades cluster, it is found \cite{martin18} that the Lyon models \cite{baraffe15} provide a more consistent account of the substellar Lithium Depletion Boundary (LDB) than older models. For brown dwarf binaries, it is found \cite{martin22} that recent evolutionary models \cite{phillips20} provide a fairly good fit to the dynamical LDB. 
Bayesian inference analysis with priors obtained from benchmark objects can be used to derive the set of properties that provides the best fit to the parameters of each SMO candidate identified by Euclid. The Bayesian model will also provide membership probability for each object in known associations, clusters, moving groups and galactic populations \cite{olivares21}.


\subsection{Probing the substellar populations of the Milky Way with Euclid}

In Figure 3, we show the distance limits for different UCD spectral types using the specifications for the Euclid surveys and comparing with Gaia. Euclid has the capability  to probe for VLM stars, UCDs, BDs and PMOs over a significant range of galactocentric distances.  The M9 class is the coolest of the M-type and it is representative of solar-metallicity VLM stars and young BDs (age $<$ 200 Myr). Gaia can detect M9 objects up to 100 pc \cite{reyle21}, while Euclid will detect them up to 1751 pc and 4398 pc with the NISP instrument in the wide and deep surveys, respectively. The Euclid VIS instrument will have a sensitivity comparable to the Vera Rubin LSST and it will overlap with it in the Fornax and South deep fields. The detection limits of the Euclid VIS and Vera Rubin LSST will be 1980 pc for M9 spectral class. The L5 spectral class is representative of the oldest solar metallicity VLM stars, middle age BDs and young PMOs (age < 50 Myr), and the T0-type is thought to be made up entirely of SMOs.  Euclid will probe for these UCDs up to distances of 624 pc and 438 pc, respectively, that include regions where there are known members in the nearest stellar associations, tidal streams and open clusters. Particularly the nearest open clusters like Coma Berenices and the Hyades, young loose associations like the Beta Pictoris, Oceanus, Tucana-Horologium and Group X moving groups, and very old streams such as Gaia-Enceladus \cite{helmi18} are widely distributed over very large areas in the sky that are partially covered by the Euclid wide survey.  

Color-color diagrams, Euclid NISP slitless spectra and reduced proper motions (from cross-matching with archive data or from follow-up observations) will be used to identify VLM stars and SMOs with metallicities below 1/10th that of the Sun using Euclid data. These objects are still very poorly known. 
Recently optical and near-infrared spectra of one of the first two extreme T subdwarfs ever discovered has been obtained with the OSIRIS and EMIR spectrographs attached to the 10.4-meter GTC, with a resolving power similar to that expected from Euclid/NISP. It is clear that there are strong changes in the spectra due to metallicity, and these effects will be detectable with Euclid/NISP. The metallicity of this object is estimated at about 1 order of magnitude below solar, the effective temperature is about 800 K, and the mass is thought to be substellar \cite{lodieu22}. Even cooler and more metal poor BDs are expected to be discovered with Euclid/NISP using spectral indicators of metallicity and temperature. Even fainter and cooler extreme subdwarfs will be identified using their peculiar colors. Proper motions could be obtained by combining Euclid and Roman data.

Radial velocities (RV) from the Euclid/NISP spectra will be obtained by cross-correlation with ultracool templates of known RV values and forward modeling methods. Even though NISP spectra have low intrinsic resolution (resolving power around 300), our simulations indicate that the highly structured absorption features in UCDs will make it possible to extract RV precision $\sim$30 km/s in L and T dwarfs with SNR$>$20 spectra, which will be essential to investigate chemo-kinetic trends in the VLM population with Euclid. For the thin disk population, we expect the substellar population to have RV absolute values $<$ 50 km/s and solar metallicity. For the thick disk substellar population, we expect a broader range of radial velocity absolute values up to about 100 km/s, and metallicity from solar to about 30 times less than that of the Sun. For the halo substellar population, we expect absolute radial velocity values $>$ 100 km/s and metallicities below one tenth that of the Sun.




f
 




\begin{figure}
\centerline{\includegraphics[width=1.0\linewidth]{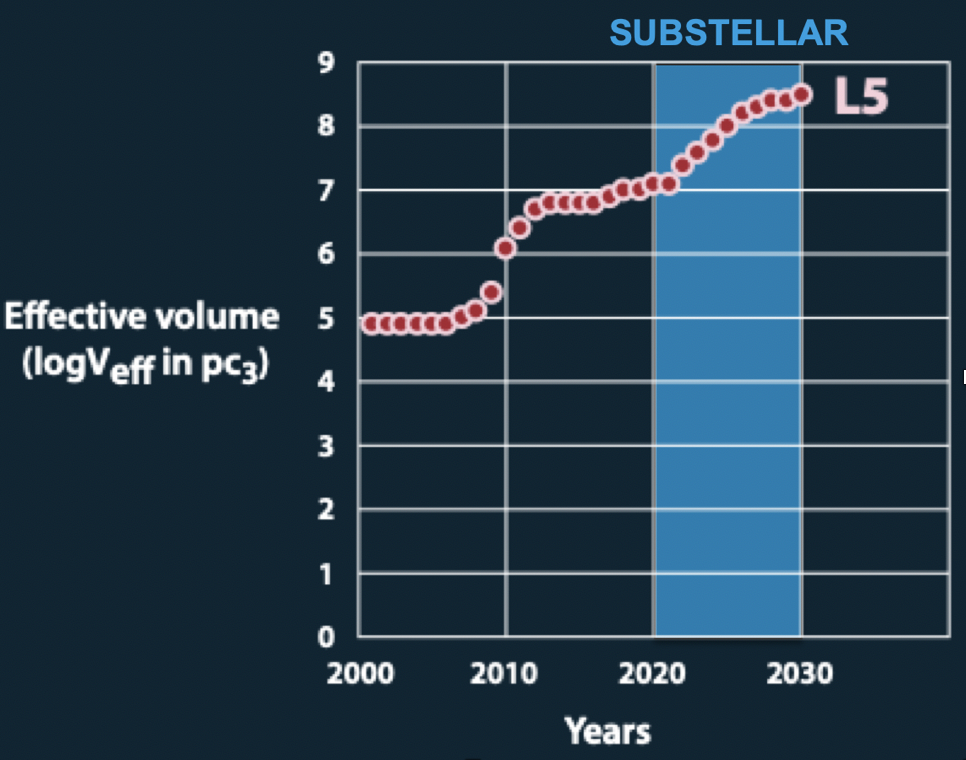}}
\caption[]{Effective volume (log Veff in pc$^3$) for detection of mid-L type dwarfs since the year 2000. The DENIS, SDSS and 2MASS surveys provided a Veff of around 10$^5$ pc$^3$ for the first decade of this century. A two order of magnitude increase was driven by the advent of the Subaru, UKIDSS and VISTA, and DES surveys during the second decade, and this decade will be dominated by another increase of almost two orders of magnitude led by the Euclid space mission and complemented by Vera Rubin Large Synoptic Telescope and the NASA Nancy Grace Roman space telescope. For mid T-type SMOs the curve looks the same but the effective volume is decreased by a factor of six. }
\label{fig:radish}
\end{figure}

\begin{figure}
\centerline{\includegraphics[width=1.0\linewidth]{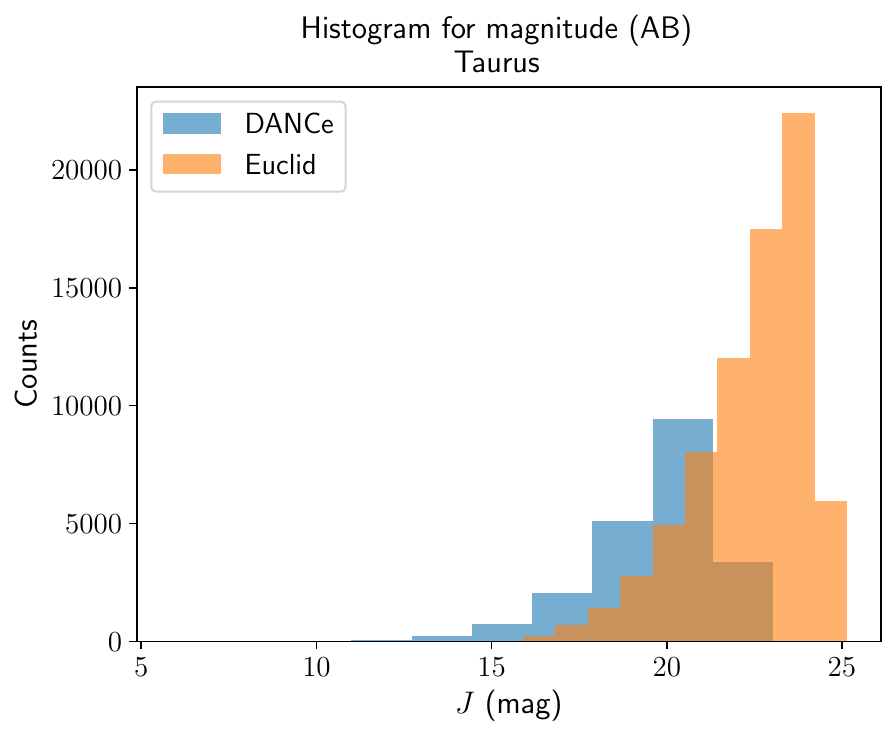}}
\caption[]{Number counts of point source J-band detections in one of the Euclid ERO002 fields compared with the number of detections in all previous surveys of the same region (DANCe catalog, H. Bouy 2023, private communication).  }
\label{fig:radish}
\end{figure}

\begin{figure}
\centerline{\includegraphics[width=1.0\linewidth]{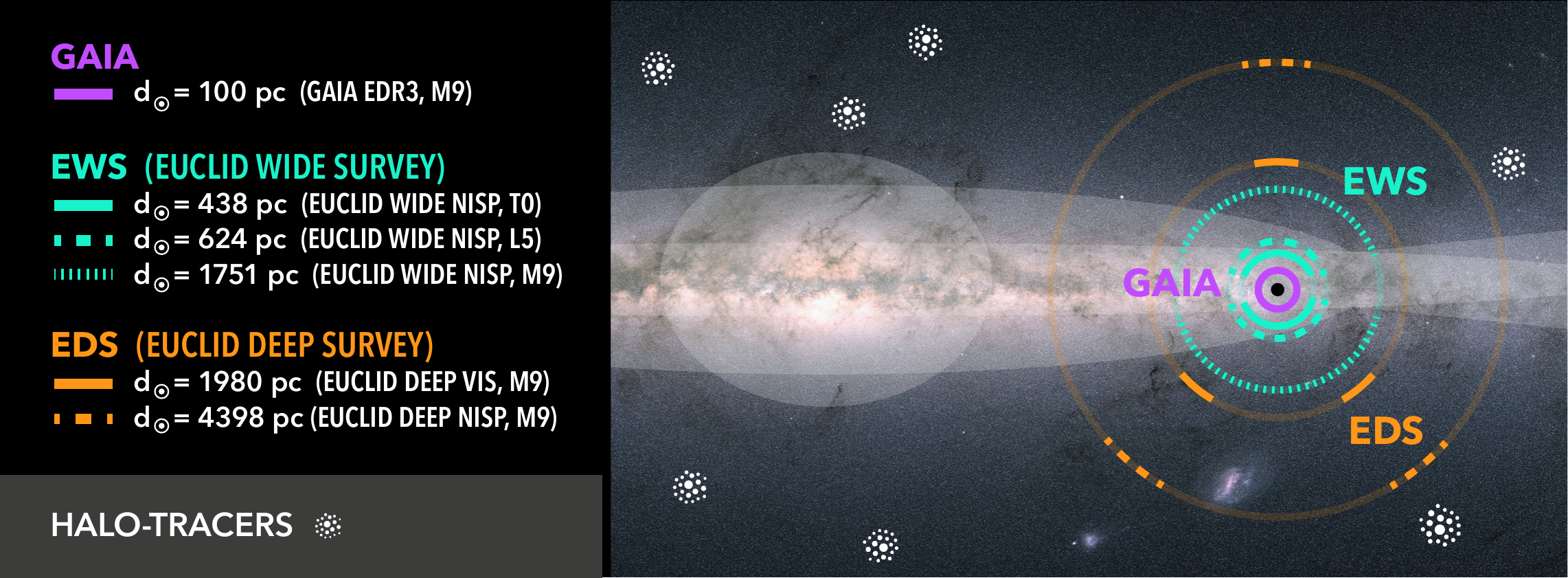}}
\caption[]{Distance limits for different UCD spectral types (M9, L5 and T0) using the specifications for the Euclid surveys \cite{scaramella22} and comparing with the Gaia completeness limit for M9-type dwarf \cite{reyle21}. Euclid has the capability  to probe for VLM stars, UCDs, BDs and PMOs over a significant range of galactocentric distances, including the halo of the Milky Way.  }
\label{fig:radish}
\end{figure}

\section*{Acknowledgments}

Funding for this paper was provided by the European Union (ERC Advanced Grant, SUBSTELLAR, project number 101054354.

\end{document}